\newif{\ifjournal}
  \journalfalse
\ifjournal
  \documentclass[]{aa}
  \usepackage{graphicx,txfonts,amssymb,natbib}
  \renewcommand{\d}{\mathrm{d}}
  \authorrunning{M. Bartelmann, M. Doran, C. Wetterich}
  \titlerunning{Non-linear structure formation with Early Dark Energy}
\else
  \documentclass{paper}
\fi

\begin{document}

\title{Non-linear Structure Formation in Cosmologies with Early Dark
  Energy}
\ifjournal
\author{Matthias Bartelmann\inst{1}, Michael Doran\inst{2} and
  Christof Wetterich\inst{2}
  \institute{$^1$ Zentrum f\"ur Astronomie, Universit\"at Heidelberg,
  Albert-\"Uberle-Str. 2, 69120 Heidelberg, Germany\\
  $^2$ Institut f\"ur Theoretische Physik, Universit\"at Heidelberg,
  Philosophenweg 16, 69120 Heidelberg, Germany}}
\else
\author{Matthias Bartelmann$^1$, Michael Doran$^2$ and Christof
  Wetterich$^2$\\
  $^1$ Zentrum f\"ur Astronomie, Universit\"at Heidelberg,
  Albert-\"Uberle-Str. 2, 69120 Heidelberg, Germany\\
  $^2$ Institut f\"ur Theoretische Physik, Universit\"at Heidelberg,
  Philosophenweg 16, 69120 Heidelberg, Germany}
\fi

\date{\today}

\newcommand{\abstext}
 {We argue that a few per cent of ``Early Dark Energy'' can be
  detected by the statistics of nonlinear structures. The presence of
  Dark Energy during linear structure formation is natural in models
  where the present tiny Dark-Energy density is related to the age of
  the Universe rather than a new fundamental small
  parameter. Generalisation of the spherical collapse model shows that
  the linear collapse parameter $\delta_\mathrm{c}$ is
  \emph{lowered}. The corresponding relative enhancement of weak
  gravitational lensing on arc-minute scales lowers the value of
  $\sigma_8$ inferred from a given lensing amplitude as compared to
  $\Lambda$CDM. In presence of Early Dark Energy, structures grow
  slower, such that for given $\sigma_8$ the number of galaxies and
  galaxy clusters is substantially increased at moderate and high
  redshift. For realistic models, the number of clusters detectable
  through their thermal Sunyaev-Zel'dovich effect at redshift unity
  and above, e.g.~with the \emph{Planck} satellite, can be an order of
  magnitude larger than for $\Lambda$CDM.}

\ifjournal
  \abstract{\abstext}
\else
  \begin{abstract}\abstext\end{abstract}
\fi

\maketitle

\section{Introduction}

According to observations of Supernovae Ia (SNe Ia) \citep{RI04.1},
the Cosmic Microwave Background (CMB)
\citep{SP03.1,RE04.2,GO03.1,RE04.3} and Large Scale Structure (LSS)
\citep{TE04.1,HA03.1}, the expansion of our Universe is accelerating
today. A plausible candidate to drive such an acceleration is Dark
Energy which may be described by an (effective, \citealt{DO02.1})
scalar field \citep{WE88.1,RA88.1,CA98.1,CA02.2}. Such scalar field
models have been shown to admit attractor solutions
\citep{WE88.1,RA88.1,LI99.1,ZL99.1} where the Dark Energy component
``tracks'' the dominant component of the cosmological fluid. It is
therefore natural to assume that Dark Energy might be non-negligible
during extended periods of the evolution of our Universe.

For a mildly varying Dark Energy fraction during the whole history of
the Universe, the present tiny Dark-Energy density can be linked to
the huge age of the Universe. Such an ``Early Dark Energy''
contribution has been shown to influence the CMB \citep{DO01.1,CA03.2}
and linear structure growth \citep{FE98.1,DO01.2}. From the CMB and
LSS data, one infers a current limit on the Dark Energy contribution
$\Omega_\mathrm{d}$ during recombination and structure formation of
$\Omega_\mathrm{d}\lesssim10\%$. Intriguingly, the effect of Early
Dark Energy on structure formation is particularly strong, because it
influences the entire structure formation period in contrast to a
cosmological constant model or any other model of Dark Energy with
$\Omega_\mathrm{d}\to0$ at earlier times.

The non-linear structure growth in Dark Energy scenarios has been
subject of several investigations (see
e.g.~\citealt{WA98.1,MA99.1,LI03.2}). In contrast to this work, we
will focus on the implications of an Early Dark Energy component. This
has two major effects. First, Early Dark Energy lowers the
\emph{linear} growth rate for structures. As a consequence, for a
given present amplitude $\sigma_8$, the amplitude of structures at
high redshift is higher as compared to $\Lambda$CDM. Second, on the
\emph{non-linear} level, there is more structure than one would
naively expect from the linear amplitude in $\Lambda$CDM. In
consequence, the abundance of collapsed objects is considerably higher
at redshifts $z\sim 1$ than in a standard $\Lambda$CDM universe. The
decisive quantity for the number of non-linear objects is the
probability for a region containing massive particles with a total
mass $M$ to collapse at a redshift $z_\mathrm{c}$. It is proportional
to
\begin{equation}
  \exp\left[
    -\frac{\delta_\mathrm{i}^2}{2\sigma_{\mathrm{i},R}^2}
  \right]\;,
\label{eq:0}
\end{equation}
where the density contrast at early times (e.g.~at matter-radiation
equality), $\delta_\mathrm{i}$, is chosen such that the region
collapses by redshift $z_\mathrm{c}$, and the variance at early times,
$\sigma_{\mathrm{i},R}$, is evaluated on a linear scale corresponding
to the mass, $R\propto(M/\rho)^{1/3}$. For a given present fluctuation
amplitude $\sigma_8$, the value of $\sigma_{\mathrm{i},R}$ is larger
in cosmologies with Early Dark Energy as compared to
$\Lambda$CDM. Since $\delta_\mathrm{i}$ depends only mildly on the
detailed cosmology, this enhancement of $\sigma_\mathrm{i}$ explains
the enhancement of non-linear structures relative to
$\Lambda$CDM. Conversely, the value of $\sigma_8$ required to explain
the number counts of non-linear objects or weak gravitational lensing
on small angular scales is smaller for models with Early Dark Energy
than for $\Lambda$CDM. Comparison of the linear fluctuation amplitude
(e.g.~from galaxy correlations or weak gravitational lensing on large
angular scales) with the power in non-linear structures could thus
provide clear evidence for the presence of Early Dark Energy.

The crucial quantity for Early Dark Energy is its average density
fraction during structure formation, $\Omega_\mathrm{d,sf}$ (see
\citealt{DO01.2} for a precise definition). As an example for the high
sensitivity of non-linear structures to $\Omega_\mathrm{d,sf}$, we
present two models with $\Omega_\mathrm{d,sf}=0.04$. We should also
mention that a possible coupling between Dark Energy and Dark Matter
could modify our results, but will be neglected in this work.

We use a particularly simple and direct parametrisation of the Dark
Energy evolution \citep{WE04.1}. The parameters are the amount of Dark
Energy today $\Omega_\mathrm{d,0}$, the equation-of-state parameter
today $w_0$, and the amount of Dark Energy at early times
$\Omega_\mathrm{d,e}$ to which it asymptotes for $z\to\infty$. One
important feature of our parametrisation is a non-vanishing
Dark-Energy contribution during recombination and structure
formation. For illustration, we pick two models at random from a
Monte-Carlo chain. We select models for which $\sigma_8$ is close to
$0.8$, the optical depth $\tau<0.2$ and the spectral index of initial
scalar perturbations is $n=0.99$ and $n=1.05$ respectively. The data
we use for the Monte-Carlo chains are CMB
\citep{SP03.1,RE04.2,GO03.1,RE04.3} and LSS data \citep{TE04.1} and
SNe Ia \citep{RI04.1}, thus both models describe current observations
well. The parameters for model (I) are:
$\Omega_\mathrm{m,0}h^2=0.146$, $\Omega_\mathrm{b}h^2=0.026$,
$h=0.67$, $n=1.05$, $\tau=0.18$, $w_0=-0.93$ and
$\Omega_\mathrm{d,e}=2\times10^{-4}$, leading to an effective Dark
Energy contribution during structure formation \citep{DO01.2} of
$\Omega_\mathrm{d,sf}=0.04$ and $\sigma_8=0.82$. Model (II) is given
by $\Omega_\mathrm{m,0}h^2=0.140$, $\Omega_\mathrm{b,0}h^2=0.023$,
$h=0.62$, $n=0.99$, $\tau=0.18$, $w_0=-0.99$ and
$\Omega_\mathrm{d,e}=8\times10^{-4}$, leading to
$\Omega_\mathrm{d,sf}=0.04$ and $\sigma_8=0.78$. We compare these
models to $\Lambda$CDM with $\Omega_\mathrm{m,0}=0.3$,
$\Omega_{\Lambda,0}=0.7$, $h=0.65$ and $\sigma_8=0.84$.

\section{Spherical collapse with Early Dark Energy}

\subsection{Solutions for early times}

The qualitative features of the effects of Early Dark Energy can be
understood analytically within the spherical collapse model. We
consider a homogeneous, spherical overdensity which expands, reaches
its maximum radius (``turn-around'') at a scale factor $a_\mathrm{ta}$
and then collapses to reach virial equilibrium. We restrict the
consideration to the matter-dominated era, i.e.~we neglect the very
slow growth of a perturbation inside the horizon during the
radiation-dominated era.

Following \cite{WA98.1}, we refer all quantities to their values at
turn-around. Let $\omega$ and $\lambda$ be the density parameters, at
turn-around, of the (Dark) Matter and the Dark Energy,
respectively. The cosmological scale factor $x$ and the radius $y$ of
the density perturbation are normalised to unity at turn-around,
\begin{equation}
  x\equiv\frac{a}{a_\mathrm{ta}}\;,\quad
  y\equiv\frac{R}{R_\mathrm{ta}}\;.
\label{eq:3}
\end{equation}
The spherical collapse of the perturbation is then described by the
two Friedmann equations
\begin{eqnarray}
  \dot x&=&\left[
    \frac{\omega}{x}+\lambda x^2g(x)+(1-\omega-\lambda)
  \right]^{1/2}\;,\label{eq:1}\\
  \ddot y&=&-\frac{\omega\zeta}{2y^2}
            -\frac{1+3w(x)}{2}\lambda g(x)y\;,
\label{eq:2}
\end{eqnarray}
where dots denote derivatives with respect to the dimension-less time
parameter
\begin{equation}
  \tau\equiv H_\mathrm{ta}\,t\;.
\label{eq:5}
\end{equation}
The parameter $\zeta$ quantifies the matter overdensity at turn-around
in a spherical volume which will collapse at a later, pre-defined
time. It is determined by solving (\ref{eq:2}) with the boundary
conditions $y=0$ at $\tau=0$ and $\dot y=0$ at turn-around, and
requiring that $y=1$ at turn-around. In other words, once the collapse
time (or redshift) is fixed when the perturbation formally collapses
to zero radius, $\zeta$ is chosen such that the maximum radius is
reached at the turn-around time, given the boundary conditions. The
function $g(x)$ quantifies the change in the Dark-Energy density with
$x$ relative to turn-around,
\begin{equation}
  g(x)=\exp\left[
    -3\int_1^x[1+w(x')]\d\ln x'
  \right]\equiv x^{-3(1+\bar w)}\;,
\label{eq:4}
\end{equation}
where $\bar w$ is a suitably averaged equation-of-state parameter. For
constant $w$, one has $\bar w=w$, but in general $w$ and $\bar w$ are
functions of $\tau$ or, equivalently, $x$.

We assume that the Dark-Energy density is always sufficiently smaller
than the matter density at early times, i.e.
\begin{equation}
  g(x)\ll x^{-3}\;,\quad x^3g(x)\ll1\quad\hbox{for}\quad x\to0\;.
\label{eq:6}
\end{equation}
Then, from (\ref{eq:1}),
\begin{equation}
  \d\tau\approx\sqrt{\frac{x}{\omega}}
  \left(1-\frac{1-\omega-\lambda}{2\omega}x\right)\d x
\label{eq:7}
\end{equation}
for $x\ll1$, and thus, at early times,
\begin{equation}
  \sqrt{\omega}\tau\approx\frac{2}{3}x^{3/2}\left[
    1-\frac{3}{10}\frac{1-\omega-\lambda}{\omega}x
  \right]\;,
\label{eq:8}
\end{equation}
without any effect of Dark Energy to second order in $x$.

Upon multiplication with $2\dot y$, Eq.~(\ref{eq:2}) becomes
\begin{equation}
  \frac{\d(\dot y^2)}{\d\tau}=\omega\zeta\frac{\d(1/y)}{\d\tau}-
  \frac{1+3w}{2}\lambda g(x)\frac{\d(y^2)}{\d\tau}\;.
\label{eq:9}
\end{equation}
Integration gives
\begin{equation}
  \dot y^2=\frac{\omega\zeta}{y}-
           \lambda\int_0^y[1+3w(x')]\,g(x')\,y'\d y'+C\;,
\label{eq:10}
\end{equation}
with the integration constant $C$ set by the requirement at
turn-around that
\begin{equation}
  \left.\dot y\right|_{y=1}=0\;,
\label{eq:11}
\end{equation}
which is satisfied if
\begin{equation}
  C=\lambda\int_0^1[1+3w(x')]\,g(x')\,y'\d y'-\omega\zeta\;.
\label{eq:12}
\end{equation}
When inserted back into (\ref{eq:10}), this yields
\begin{equation}
  \dot y^2=\omega\zeta\left(\frac{1}{y}-1\right)+
  \lambda\int_y^1[1+3w(x')]\,g(x')\,y'\d y'\;.
\label{eq:13}
\end{equation}

We will have to study (\ref{eq:13}) at very early times, i.e.~for
$x\to0$ and $y\to0$. In that limit, the integral (\ref{eq:13}) will
either converge or diverge as $x$ and $y$ approach zero. Suppose the
integral converges,
\begin{equation}
  \lim_{y\to0}\int_y^1(1+3w)g(x)y\d y\equiv I<\infty\;.
\label{eq:14}
\end{equation}
Then, for $y\to0$, Eq.~(\ref{eq:13}) can be approximated by
\begin{equation}
  \frac{\sqrt{y}\d y}{\sqrt{1-Ay}}=\sqrt{\omega\zeta}\d\tau\;,\quad
  A\equiv1-\frac{\lambda I}{\omega\zeta}\;,
\label{eq:15}
\end{equation}
which can be integrated after expanding the denominator into a Taylor
series,
\begin{equation}
  \sqrt{\omega\zeta}\tau\approx
  \int_0^y\sqrt{y}\left(1+\frac{Ay}{2}\right)=
  \frac{2}{3}y^{3/2}\left(1+\frac{3A}{10}y\right)\;.
\label{eq:16}
\end{equation}
For example, if the Dark Energy is described by a cosmological
constant, $w=-1$ and $g(x)=1$, thus $I=-1$ and
$A=1+\lambda/\omega\zeta$.

The integral $I$ defined in (\ref{eq:14}) will diverge if $g(x)$
increases steeply as $x\to0$. Then, it will be dominated by the
behaviour of its integrand for $x\to0$. Let
\begin{equation}
  \lim_{x\to0}w(x)\equiv w_\mathrm{ini}\;,\quad
  \lim_{x\to0}g(x)\equiv\gamma x^{-3(1+w_\mathrm{ini})}
\label{eq:17}
\end{equation}
with a constant $\gamma\ll1$, and we further assume $y$ to be
proportional to $x$ at early times, $y=\alpha x$, which will be
justified in hindsight. Then,
\begin{eqnarray}
  I&\approx&(1+3w_\mathrm{ini})\,\alpha^2\gamma\,
  \lim_{x\to0}\int_x^1x'^{-3(1+w_\mathrm{ini})}\,x'\d x'\nonumber\\
  &\approx&\alpha^2\gamma\,\lim_{x\to0}\left(\left.
    x'^{-1-3w_\mathrm{ini}}\right|_x^1\right)\nonumber\\
  &\approx&
  \frac{\gamma\alpha^{3(1+w_\mathrm{ini})}}{y^{1+3w_\mathrm{ini}}}\;.
\label{eq:18}
\end{eqnarray}
$I$ is thus expected to diverge if $w_\mathrm{ini}\ge-1/3$. With
(\ref{eq:13}), this yields for early times
\begin{equation}
  \dot y=\left[\frac{\omega\zeta}{y}-\omega\zeta+
    \lambda\gamma
    \frac{\alpha^{3(1+w_\mathrm{ini})}}{y^{1+3w_\mathrm{ini}}}
  \right]^{1/2}\;.
\label{eq:19}
\end{equation}

According to (\ref{eq:6}), $w_\mathrm{ini}\le0$, because otherwise the
Dark-Energy density would grow above the matter density. Let
$w_\mathrm{ini}=0$ first, then (\ref{eq:19}) implies
\begin{eqnarray}
  \dot y&=&\left[
    \frac{\omega\zeta+\lambda\gamma\alpha^3}{y}-\omega\zeta
  \right]^{1/2}\nonumber\\
  &\approx&
  \sqrt{\frac{\omega\zeta+\lambda\gamma\alpha^3}{y}}\,\left(
    1-\frac{\omega\zeta}{2(\omega\zeta+\lambda\gamma\alpha^3)}y
  \right)\;,
\label{eq:20}
\end{eqnarray}
which can be integrated to give
\begin{equation}
  \sqrt{\omega\zeta}\tau\approx
  \frac{2}{3}y^{3/2}\,\sqrt{B}\left(
    1+\frac{3}{10}By
  \right)\;,\quad
  B\equiv\frac{\omega\zeta}{\omega\zeta+\lambda\gamma\alpha^3}\;.
\label{eq:21}
\end{equation}
If $-1/3<w_\mathrm{ini}<0$, we can replace $B$ by
\begin{equation}
  B=\frac{\omega\zeta}
    {\omega\zeta+\lambda\gamma\alpha^{3(1+w_\mathrm{ini})}
    y^{-3w_\mathrm{ini}}}\approx
  1-\frac{\lambda\gamma}{\omega\zeta}\alpha^{3(1+w_\mathrm{ini})}
    y^{-3w_\mathrm{ini}}
\label{eq:22}
\end{equation}
to account for the gentle change in the coefficient $B$ with $y$.

Equating $\sqrt{\zeta}$ times (\ref{eq:8}) to (\ref{eq:16}) and
squaring yields
\begin{equation}
  \zeta x^3\left(1-\frac{3}{5}\frac{1-\omega-\lambda}{\omega}x\right)
  \approx y^3\,\left(1+\frac{3A}{5}y\right)\;,
\label{eq:23}
\end{equation}
while the same procedure applied to (\ref{eq:21}) implies
\begin{equation}
  \zeta x^3\left(1-\frac{3}{5}\frac{1-\omega-\lambda}{\omega}x\right)
  \approx y^3B\,\left(1+\frac{3B}{5}y\right)\;.
\label{eq:24}
\end{equation}

Now, $B\approx1$ to lowest order because $\gamma\ll1$ and $y\ll1$ at
early times, thus
\begin{equation}
  y\approx\zeta^{1/3}x\;,\quad\alpha=\zeta^{1/3}
\label{eq:25}
\end{equation}
to lowest order, and the earlier assumption $y\approx\alpha x$ is
verified. The definition of $B$ in (\ref{eq:21}) and (\ref{eq:22})
then simplifies to
\begin{equation}
  B\approx
  1-\frac{\lambda\gamma\zeta^{w_\mathrm{ini}}
  y^{-3w_\mathrm{ini}}}{\omega}
\label{eq:26}
\end{equation}

In order to illustrate the procedure described above, let us assume
$w(a)=-a$, or $w(x)=-a_\mathrm{ta}x$, for which $g(x)$ from
(\ref{eq:4}) becomes
\begin{equation}
  g(x)=\exp\left[-3\int_1^x(1-a_\mathrm{ta}x')\,\d\ln x'\right]=
  \frac{e^{3a_\mathrm{ta}(x-1)}}{x^3}\;.
\label{eq:27}
\end{equation}
This yields
\begin{eqnarray}
  \int_y^1[1+3w(x')]\,g(x')\,y'\d y'&\approx&\alpha^2\left[
    \frac{e^{3a_\mathrm{ta}(x-1)}}{x}-1
  \right]\nonumber\\
  &\approx&\frac{\alpha^2e^{-3a_\mathrm{ta}}}{x}\approx
  \frac{\alpha^3e^{-3a_\mathrm{ta}}}{y}\;.
\label{eq:28}
\end{eqnarray}
Following (\ref{eq:17}), we have
\begin{equation}
  w_\mathrm{ini}=0\;,\quad\gamma=e^{-3a_\mathrm{ta}}\;,
\label{eq:29}
\end{equation}
confirming that $\gamma\ll1$ for typical values of
$a_\mathrm{ta}\lesssim1$. Thus, (\ref{eq:28}) agrees with the more
general result (\ref{eq:18}).

\subsection{Overdensity and linear density contrast}

Since the background matter density changes in proportion to $x^{-3}$,
while the matter density inside the perturbation is proportional to
$y^{-3}$, the overdensity inside the perturbation obeys
\begin{equation}
  \Delta=\frac{\zeta x^3}{y^3}\;,
\label{eq:30}
\end{equation}
which, according to (\ref{eq:23}) and (\ref{eq:24}), is
\begin{equation}
  \Delta\approx\left\{
  \begin{array}{ll}
    \displaystyle
    1+\frac{3A}{5}y+\frac{3}{5}\frac{1-\omega-\lambda}{\omega}x
    & \hbox{for}\quad w_\mathrm{ini}<-1/3 \\[12pt]
    \displaystyle
    1+\frac{3B}{5}y+\frac{3}{5}\frac{1-\omega-\lambda}{\omega}x
    & \hbox{else} \\
  \end{array}\right.
\label{eq:31}
\end{equation}
at early times. The two solutions join at $w_\mathrm{ini}=-1/3$, as
they should. For $w_\mathrm{ini}=-1/3$, $I=0$ because
$(1+3w_\mathrm{ini})=0$ in Eq.~(\ref{eq:18}), thus $B=1$. Also,
$A\to1$ for $w\to-1/3$ according to (\ref{eq:14}) and (\ref{eq:15}).

The overdensity $\Delta_\mathrm{v}$ within virialised objects should
only very weakly be affected by the (Early) Dark Energy because Dark
Matter dominates the virialisation process. We verify this as
described in Appendix~\ref{app:1}.

The \emph{linear} density \emph{contrast} $\delta$ can be used to
relate the non-linear overdensity $\Delta$ with the density that would
result from linear evolution of the same initial perturbation within
linear evolution. With the growth factor $D_+(x)$ at a given time or
normalised scale factor $x$ one has at the collapse time
$x_\mathrm{c}>1$
\begin{equation}
  \delta_\mathrm{c}=\lim_{x\to0}\left[
    \frac{D_+(x_\mathrm{c})}{D_+(x)}[\Delta(x)-1]
  \right]\;,
\label{eq:34}
\end{equation}
The linear density contrast at collapse time $\delta_\mathrm{c}$ is a
crucial ingredient for the Press-Schechter and related mass functions
\citep{PR74.1,SH99.1}. We shall explain below why $\delta_\mathrm{c}$
is substantially smaller for Early Dark-Energy models as compared to
$\Lambda$CDM. This finding is a central ingredient for our results.

\subsection{Results for $\delta_\mathrm{c}$}

\begin{figure}[ht]
  \includegraphics[width=\hsize]{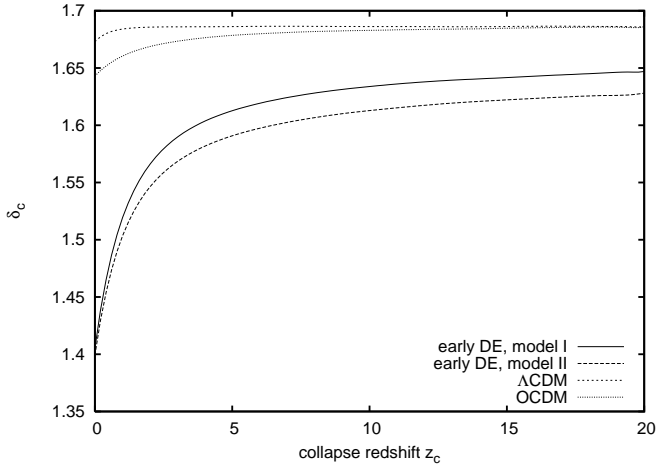}
\caption{The linear density contrast at collapse, $\delta_\mathrm{c}$
  is plotted as a function of the collapse redshift $z_\mathrm{c}$ for
  two models with early Dark Energy, and for $\Lambda$CDM and OCDM for
  comparison. While the $\Lambda$CDM and the OCDM models show very
  similar behaviour, with $\delta_\mathrm{c}$ near the ``canonical''
  value of $1.686$, the Early-DE models fall substantially below, in
  particular for low collapse redshifts.}
\label{fig:2}
\end{figure}

We show in Fig.~\ref{fig:2} the linear critical overdensity
$\delta_\mathrm{c}$ as a function of $z_\mathrm{c}$. While the results
range around $\delta_\mathrm{c}=3/5(3\pi/2)^{2/3}\approx1.686$ for
$\Lambda$CDM and OCDM, quite independent of the collapse redshift,
they fall below for the Early-DE models, dropping even to
$\delta_\mathrm{c}\approx1.4$ for $z_\mathrm{c}=0$.

\begin{figure}[ht]
  \includegraphics[width=\hsize]{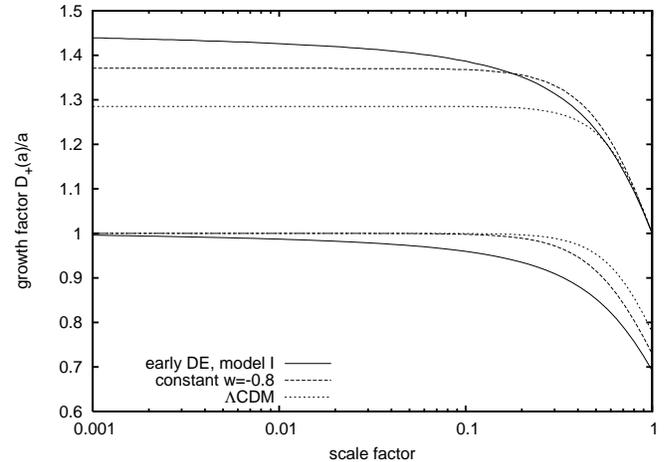}
\caption{The linear growth factor divided by the scale factor,
  $D_+(a)/a$, is shown here as a function of $a$ for an Early-DE
  model, a model with constant $w=-0.8$, and for the $\Lambda$CDM
  model. The lower curves are normalised to unity at early times,
  i.e.~they illustrate the growth of perturbations starting from the
  same density contrast in the two cosmologies. The upper curves are
  normalised to unity at present ($a=1$), illustrating the growth of
  structures reaching the same level today in both
  cosmologies. Starting from the same amplitude, the growth in the
  Early-DE model falls behind that in $\Lambda$CDM, largely causing
  the reduction in the overdensity parameter
  $\delta_\mathrm{c}$. Structures have to grow earlier in Early-DE
  models than in $\Lambda$CDM to reach the same fluctuation amplitude
  today.}
\label{fig:3}
\end{figure}

We return to (\ref{eq:34}) in order to understand why
$\delta_\mathrm{c}$ is lower for Early-DE than for, e.g., the
$\Lambda$CDM model. It turns out numerically that the overdensity
$(\Delta-1)$ is not changed much in the Early-DE compared to the
$\Lambda$CDM model. For a collapse redshift of $z_\mathrm{c}=0$, for
instance, $(\Delta-1)/x$ is lower by only $4\%$ in the Early-DE
than in the $\Lambda$CDM model. The main difference, however, is
caused by the growth-factor ratio in (\ref{eq:34}), as shown in
Fig.~\ref{fig:3}.

The curves show the growth factor divided by the scale factor,
$D_+(a)/a$, as a function of $a$, for one of the Early-DE models and
for $\Lambda$CDM. Both lower curves are normalised such that they
start from unity at early times, as it should be according to
(\ref{eq:34}). Obviously, the linear growth in Early-DE falls behind
that in $\Lambda$CDM, starting from an overdensity of the same
amplitude. The reason is that the expansion rate in the Early-DE
models is higher at early times than in $\Lambda$CDM, which increases
the friction term $(\dot a/a)$ in the equation
\begin{equation}
  \ddot\delta+2\frac{\dot a}{a}\dot\delta-4\pi G\rho\delta=0
\label{eq:41}
\end{equation}
governing the growth factor. As we shall see below, the reduced linear
overdensity $\delta_\mathrm{c}$ expected in Early-DE models has
pronounced consequences for nonlinear structure formation.

The two upper curves in Fig.~\ref{fig:3} show the growth factor
normalised to unity at present, i.e.~reflecting the evolution back in
time of structures reaching the same amplitude today. They show that
structures need to grow earlier in Early-DE models than in
$\Lambda$CDM models to reach the same level at the present time.

\section{Cosmological consequences}

\subsection{Distances and age}

\begin{figure}[ht]
  \includegraphics[width=\hsize]{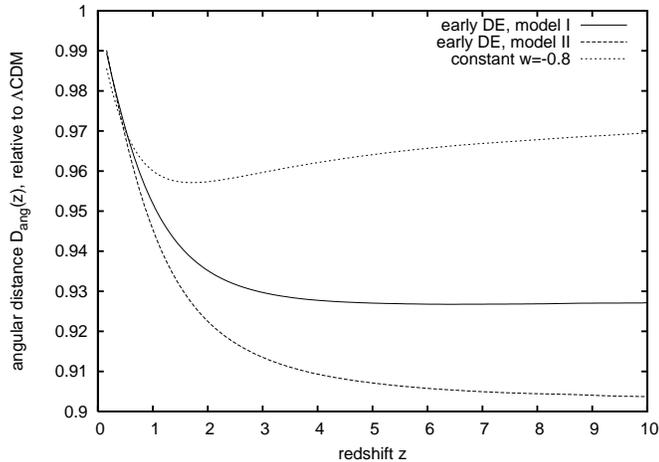}
\caption{The angular-diameter distance as a function of redshift is
  shown for two Early-DE models relative to $\Lambda$CDM. A model with
  constant $w=-0.8$ is also shown for comparison.}
\label{fig:4}
\end{figure}

The influence of Early Dark Energy on global geometrical properties of
the Universe is illustrated in Fig.~\ref{fig:4}, which shows the
angular-diameter distance $D_\mathrm{ang}(z)$ in units of $cH_0^{-1}$
as a function of redshift relative to $\Lambda$CDM. A model with
constant $w=-0.8$ is shown for comparison. Note that this curve also
gives the ratio of luminosity distances $D_\mathrm{lum}(z)$ between
Dark-Energy models and $\Lambda$CDM because $D_\mathrm{lum}(z)$ and
$D_\mathrm{ang}(z)$ only differ by the ratio of scale factors between
emission and observation.

\begin{figure}[ht]
  \includegraphics[width=\hsize]{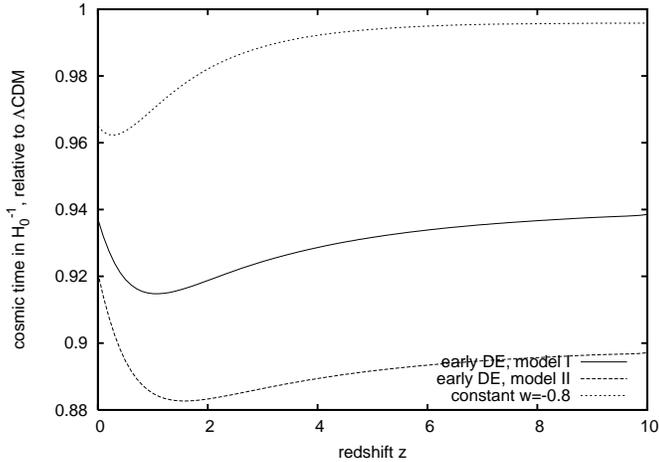}
\caption{Cosmic time in units of $H_0^{-1}$ as a function of redshift,
  relative to $\Lambda$CDM. Three curves are given, two for the
  Early-DE models and one for a constant $w=-0.8$ for comparison. For
  fixed Hubble constant $H_0$, Early Dark Energy makes the universe
  younger today by $(5-10)\%$ compared to $\Lambda$CDM, while a fixed
  present age would reduce $H_0$.}
\label{fig:5}
\end{figure}

As the figure shows, distance measures are changed only moderately, by
$\lesssim8\%$ for source redshifts below $z=2$. The cosmic time is
changed by a larger amount, as can be seen in Fig.~\ref{fig:5}. It
shows the cosmic time as a function of redshift in units of
$H_0^{-1}$, again relative to $\Lambda$CDM.

The increased expansion rate of the Universe compared to $\Lambda$CDM,
in particular at early times, reduces the age of the Universe by
approximately $(5-10)\%$ at low redshifts. The effect of Dark Energy
with constant $w=-0.8$ is substantially less pronounced.

\subsection{The mass function}

\begin{figure}[ht]
  \includegraphics[width=\hsize]{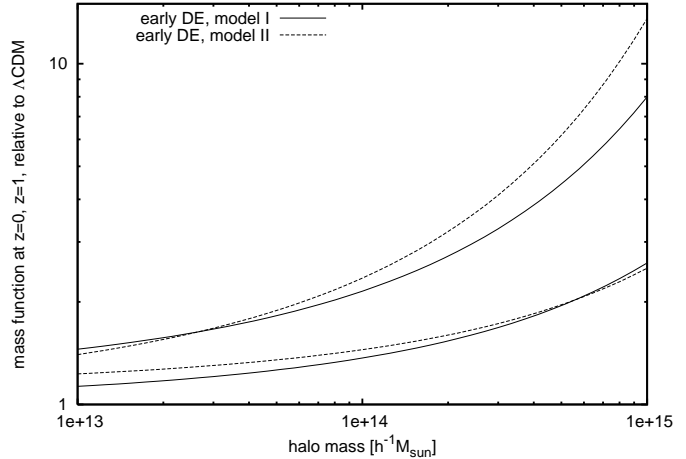}
\caption{Ratio between the Dark-Matter halo mass function between
  $10^{13}$ and $10^{15}\,h^{-1}\,M_\odot$ in two Early-Dark Energy
  models compared to $\Lambda$CDM. Two pairs of curves are given, the
  lower for $z=0$, the upper for $z=1$. The ratio increases
  monotonically with increasing mass. While massive clusters today are
  $1.5-2.5$ times more abundant in CMB-normalised Early-DE compared to
  the $\Lambda$CDM model, their number is expected to be up to an
  order of magnitude higher at redshift $z=1$ in Early-DE models.}
\label{fig:6}
\end{figure}

The reduced linear density contrast $\delta_\mathrm{c}$ necessary for
spherical collapse has a pronounced influence on the mass function of
Dark-Matter halos. The Press-Schechter mass function for example can
be written as \citep{PR74.1}
\begin{eqnarray}
  N(M,z)\d M&=&\sqrt{\frac{2}{\pi}}\,
  \frac{\rho_0\delta_\mathrm{c}}{\sigma_RD_+(z)}\,
  \frac{\d\ln\sigma_R}{\d M}\nonumber\\
  &\times&
  \exp\left(-\frac{\delta_\mathrm{c}^2}{2\sigma_R^2D_+^2(z)}\right)\,
  \frac{\d M}{M}\;,
\label{eq:42}
\end{eqnarray}
showing that $\delta_\mathrm{c}^2$ is compared in the argument of the
exponential function to the variance $\sigma_R^2D_+^2(z)$ of the
Dark-Matter density field smoothed on scales $R$ corresponding to the
mass $M$, with $D_+(z=0)=1$. Linear quantities occur in the
exponential of the Press-Schechter mass function because it predicts
the distribution of the later non-linear fluctuations from their
linear distribution at early times. Even small reductions of
$\delta_\mathrm{c}$, as they are illustrated in Fig.~\ref{fig:2}, lead
to a noticeable increase in the mass function for a fixed
normalisation parameter $\sigma_8$ of the Dark-Matter power
spectrum. The increase of the mass function in the two Early Dark
Energy models compared to $\Lambda$CDM is illustrated in
Fig.~\ref{fig:6}.

\begin{figure}[ht]
  \includegraphics[width=\hsize]{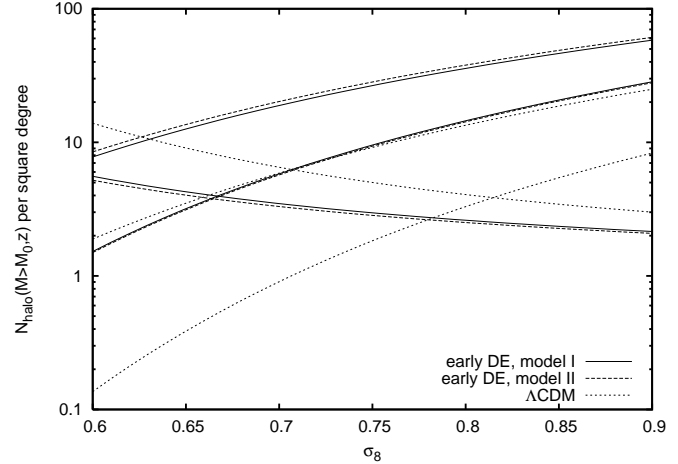}
\caption{\emph{Rising curves}: The number of halos per square degree
  with masses $M\ge10^{14}\,h^{-1}\,M_\odot$ is shown as a function of
  the normalisation parameter $\sigma_8$. The upper and lower curves
  show the numbers of halos with redshifts $z\ge0$ and $z\ge1$,
  respectively. At $\sigma_8\sim0.7$, the Early-DE models have
  $\sim25$ cluster-sized halos with $z\ge0$ per square degree, for
  which the $\Lambda$CDM model needs $\sigma_8\sim0.9$. The
  \emph{falling curves} show the ratios of halo numbers with mass
  $M\ge10^{14}\,h^{-1}\,M_\odot$ above redshifts zero and unity for
  the two models of Early Dark Energy and for $\Lambda$CDM as a
  function of the normalisation parameter $\sigma_8$. At fixed
  $\sigma_8\sim0.8$, the number of cluster-sized halos grows by
  approximately twice as much in the $\Lambda$CDM compared to the
  Early-DE models.}
\label{fig:7a}
\end{figure}

The figure shows the mass function proposed by \cite{SH99.1}, which
assumes ellipsoidal rather than spherical halo collapse and reproduces
the halo abundances in simulations significantly better than the
original Press-Schechter mass function. As Fig.~\ref{fig:6} shows,
Early Dark Energy has a pronounced effect on the mass function. While
halos with masses $\lesssim10^{13}\,h^{-1}\,M_\odot$ have
approximately equal abundances as in $\Lambda$CDM, massive clusters
with $M\gtrsim5\times10^{14}\,h^{-1}\,M_\odot$ are of order $1.5-2.5$
times more abundant at the present epoch, provided the models are
normalised according to the CMB temperature fluctuations. At redshift
unity, however, the mass functions differ by a factors of $\sim(5-10)$
for massive clusters. Therefore, even if the models were normalised
such as to get closer to the cluster abundance expected in
$\Lambda$CDM today, they still predicted a higher number density of
high-redshift clusters. As Fig.~\ref{fig:7a} shows in more detail, the
normalisation parameter would have to be lowered to
$\sigma_8\approx0.69$ for the Early-DE models to reproduce the halo
abundance at $z=0$ of a $\Lambda$CDM model with $\sigma=0.84$.

\begin{figure}[ht]
  \includegraphics[width=\hsize]{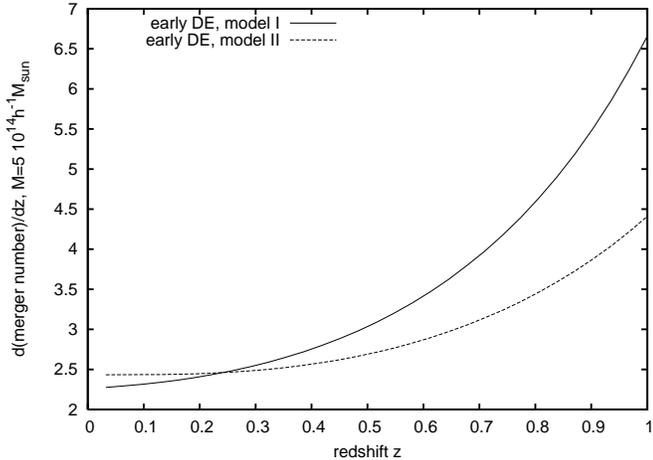}
\caption{The number per unit redshift of cluster-sized halos with mass
  $M=5\times10^{14}\,h^{-1}\,M_\odot$ is shown here which have
  undergone a major merger with another halo of mass between $M/4$ and
  $M$ within redshifts $z$ and $z+0.2$. Both curves are normalised to
  the merger number in $\Lambda$CDM. Reflecting the earlier growth of
  the mass function in Early-DE models compared to $\Lambda$CDM, major
  mergers of massive halos are about a factor of $\approx5$ more
  frequent at redshifts near unity.}
\label{fig:7}
\end{figure}

The earlier development of the mass function naturally leads to
enhanced dynamical activity of Dark-Matter halos, in particular at the
highest cluster masses. The extended Press-Schechter formalism
\citep{BO91.1} allows the calculation of the merger probability,
\begin{equation}
  \frac{\d^2p}{\d\ln\Delta M\d t}(M,z)\;,
\label{eq:43}
\end{equation}
i.e.~the probability for a halo of mass $M$ to merge within time $t$
at redshift $z$ with another halo of mass $\Delta M$
\citep{LA93.1,LA94.1}. Multiplying (\ref{eq:43}) with the mass function
and the differential cosmic volume $\d V$ corresponding to a redshift
interval $\d z$, and integrating over a mass range from $M_1$ to $M_2$
and over the time $t$ corresponding to a redshift interval $z-\delta
z$ to $z$ yields the total number of mergers between a halo of mass
$M$ and another halo of mass between $M_1$ and $M_2$, within the time
interval from $t(z+\delta z)$ and $t(z)$,
\begin{eqnarray}
  \delta N(M,z)&=&N(M,z)\left|\frac{\d V}{\d z}\right|\,
  \int_{\ln M_1}^{\ln M_2}\d\ln M\nonumber\\
  &\times&\int_{t(z+\delta z)}^{t(z)}\d t\,
  \frac{\d^2p(M,z)}{\d\ln M\d t}\;.
\label{eq:44}
\end{eqnarray}
We set $M=5\times10^{14}\,h^{-1}\,M_\odot$, $M_1=M/4$, $M_2=M$, and
$\delta z=0.2$. In other words, we compute using expression
(\ref{eq:44}) the number per unit redshift of cluster-sized
Dark-Matter halos at redshift $z$ that have undergone a major merger
within $z$ and $z+0.2$. Figure~\ref{fig:7} shows the ratio between the
merger number in the two Early-DE models relative to $\Lambda$CDM.

The two curves in the figure show that major mergers at moderate and
high redshifts of cluster-sized halos are substantially more frequent
in the two Early Dark-Energy models compared to $\Lambda$CDM. This
reflects our previous result that the halo mass function grows earlier
in the Early-DE models, leading to enhanced merger activity at earlier
times.

Of course, these results on the abundance of Dark-Matter halos depend
sensitively on the normalisation of the power spectrum, as the rising
curves in Fig.~\ref{fig:7a} illustrate. Much less dependent on the
amplitude of the power spectrum is the expected ratio of halo numbers
above two different redshifts. The falling curves in Fig.~\ref{fig:7a}
give examples.

The curves show for three cosmological models in dependence of the
normalisation parameter $\sigma_8$ by how much the number of
cluster-sized halos with mass $M\ge10^{14}\,h^{-1}\,M_\odot$ grows
between redshifts unity and zero. For $\sigma_8=0.8$ in $\Lambda$CDM,
there are about four times more clusters today than there were at
redshift unity, and about two times more in the Early-DE models. This
corroborates that clusters are more abundant at high redshifts in the
Early-DE models compared to $\Lambda$CDM. Since CDM is a hierarchical
model of structure formation, a similar effect appears for lower-mass
halos at higher redshift. In the Early-DE models, for instance, the
number of galaxy-sized halos evolves much less than in $\Lambda$CDM
above redshifts 2 and beyond.

\subsection{Halo properties}

Earlier work has found that Dark-Matter halos tend to be more
concentrated if they form earlier, where the concentration $c$ is the
ratio between the halo's virial and scale radii
\citep{NA97.1,BU01.1,EK01.1}. While the central slope of Dark-Matter
density profiles in halos is still under some debate
(cf.~\cite{PO03.1}), numerical simulations consistently show that
halos in (Cold) Dark Matter have a steep density profile outside, and
a flat density profile inside the scale radius. The definition of the
virial radius changes in the literature, depending on whether the mean
density enclosed by the virial radius is supposed to be 200 or another
factor times the mean or the critical cosmic density. Different
recipes have also been given for the statistical relation between the
virial mass and the concentration of halos. They were tested against
numerical results on Dark-Energy models by \cite{DO03.2} who found
that the algorithm described by \cite{EK01.1} worked for models with
constant $w\ne-1$ without adaptation.

\begin{figure}[ht]
  \includegraphics[width=\hsize]{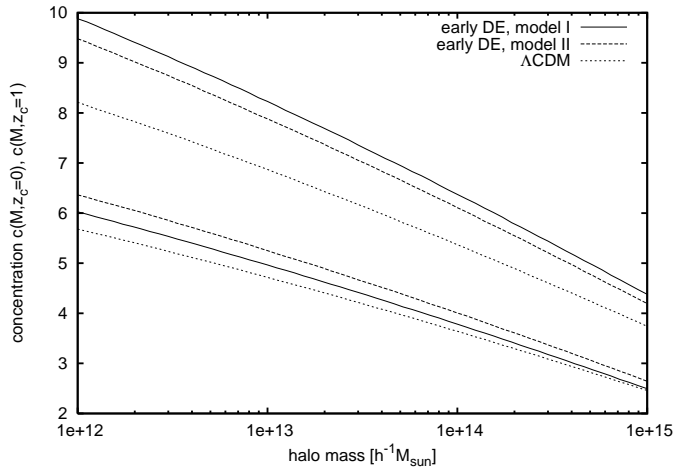}
\caption{Halo concentrations as functions of mass, according to Eke et
  al.'s algorithm, for redshifts unity (lower curves) and zero (upper
  curves). Results for the two Early-DE models and for $\Lambda$CDM
  are given. Halos in Early-DE models are more concentrated than in
  $\Lambda$CDM, and the difference increases for smaller masses.}
\label{fig:8}
\end{figure}

All algorithms have in common that they have to define a collapse
redshift $z_\mathrm{c}$ for a halo of given mass $M$ at redshift $z$,
and a ratio between the central density of the halo and the mean or
critical cosmic background density at $z_\mathrm{c}$. \cite{EK01.1}
define the collapse redshift by requiring that, at $z_\mathrm{c}$, the
amplitude of the power spectrum at the mass scale of the halo reaches
a given fixed value. We show in Fig.~\ref{fig:8} the expected
concentration according to Eke et al.'s prescription as a function of
halo mass at redshifts zero (upper curves) and unity (lower curves).

In agreement with earlier studies \citep{BA02.1,KL03.1,DO03.2,WE03.1},
halos in cosmologies with dynamical Dark Energy tend to be more
concentrated than in $\Lambda$CDM, reflecting their earlier growth on
a cosmological background with higher mean density. For galaxy-sized
halos in the Early-DE models, the concentration increases by
$\approx20\%$ at redshift zero.

\subsection{Gravitational lensing}

The efficiency of gravitational lensing by isolated lenses such as
galaxies or galaxy clusters is given by the critical surface-mass
density
\begin{equation}
  \Sigma_\mathrm{cr}=\frac{c^2}{4\pi G}\,
  \frac{D_\mathrm{s}}{D_\mathrm{l}D_\mathrm{ls}}\;,
\label{eq:45}
\end{equation}
which contains the angular-diameter distances $D_\mathrm{l,s,ls}$ from
the observer to the lens, the source, and from the lens to the source,
respectively (e.g.~\cite{SC92.1}). The effective lensing distance
$D_\mathrm{eff}\equiv D_\mathrm{l}D_\mathrm{ls}/D_\mathrm{s}$ thus
measures the geometrical efficiency of a given mass distribution.

\begin{figure}[ht]
  \includegraphics[width=\hsize]{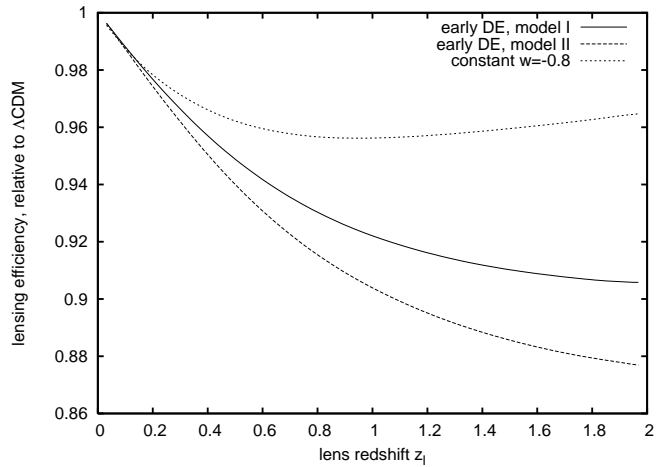}
\caption{The effective lensing distance is plotted here for the two
  Early-DE models, normalised by the result in $\Lambda$CDM. The curve
  for constant $w=-0.8$ is shown for comparison. Sources are assumed
  at redshift $z_\mathrm{s}=2$. The effective lensing distance is
  lower by $\approx5\%$ in Early-DE than in $\Lambda$CDM models at
  typical lens redshifts near $z=0.4$, and by $\lesssim10\%$ at
  redshifts near unity.}
\label{fig:9}
\end{figure}

Dark Energy reduces the effective lensing distance compared to
$\Lambda$CDM. For sources at redshift $z_\mathrm{s}=2$, as assumed for
the figure, identical mass distributions are $\approx5\%$ less
efficient at $z_\mathrm{l}\approx0.4$, and $\approx10\%$ less
efficient at $z_\mathrm{l}\approx1.0$, in Early-DE than in
$\Lambda$CDM models. Compared to the increased concentration of the
Dark-Matter halos, this effect is less important for strong lensing by
galaxies, groups or clusters \citep{ME04.1}.

\begin{figure}[ht]
  \includegraphics[width=\hsize]{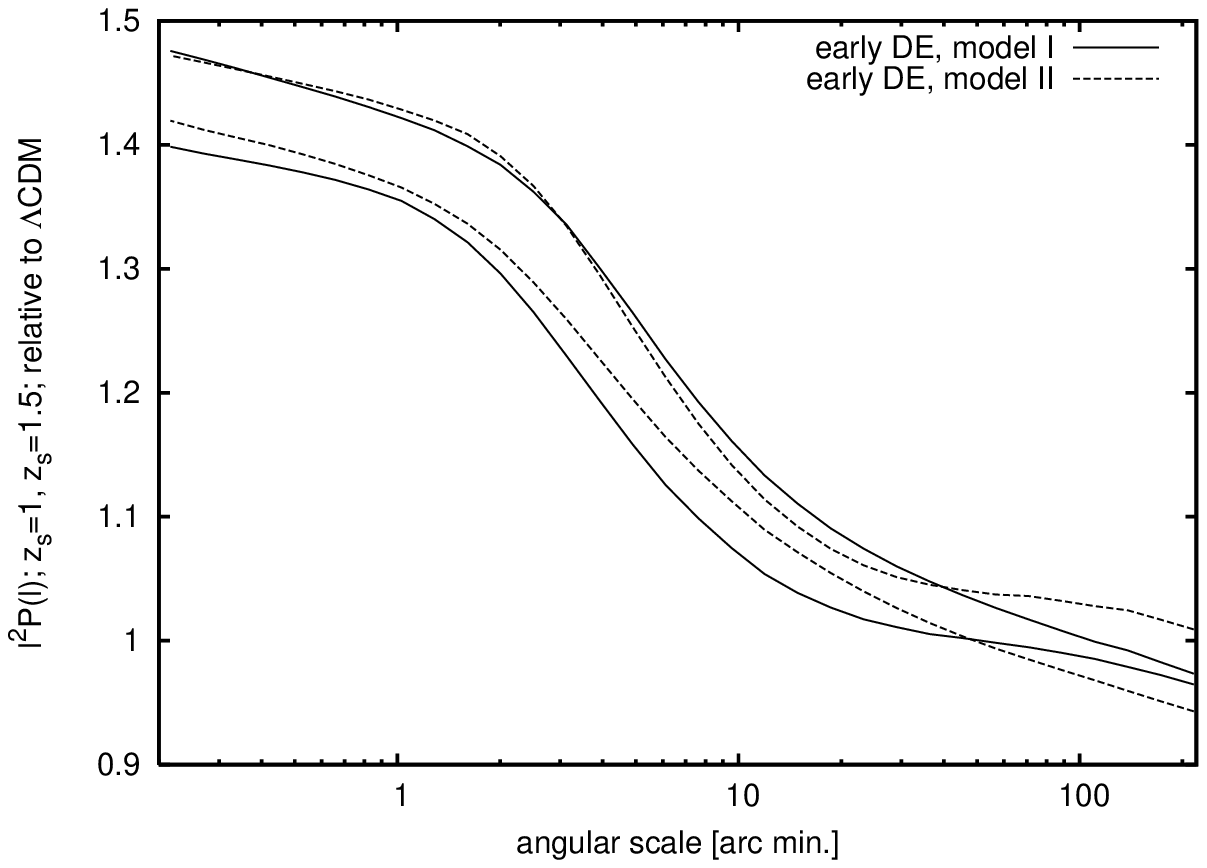}
\caption{The weak-lensing power, expressed by means of the convergence
  power spectrum as $l^2P_\kappa(l)$, is plotted as a function of
  angular scale for the two Early-DE models for the two different
  source redshifts $z_\mathrm{s}=1$ and $1.5$. The curves are
  normalised by the expectation in $\Lambda$CDM.}
\label{fig:10}
\end{figure}

In sufficient approximation, weak gravitational lensing by large-scale
structures can be described by the power spectrum $P_\kappa(l)$ of the
effective convergence $\kappa$ (see \cite{BA01.1} for a
review). Earlier structure growth also affects $P_\kappa$, as shown in
Fig.~\ref{fig:10}.

The weak-lensing power is illustrated there plotting $l^2P_\kappa(l)$
as a function of angular scale for the two Early-DE models, relative
to the $\Lambda$CDM model. Two pairs of curves are given for the two
source redshifts $z_\mathrm{s}=1$ and $1.5$. The curves tend to unity
at angular scales above $10$ arc minutes, indicating that weak lensing
will not be modified compared to $\Lambda$CDM on large angular
scales. On small angular scales $\lesssim1'$ however, the weak-lensing
power in the Early-DE models exceeds that in the $\Lambda$CDM model by
$\sim40\%$ for $z_\mathrm{s}=1$ and by $\sim55\%$ for
$z_\mathrm{s}=1.5$. This would imply that $\sigma_8$ as derived from
weak-lensing measurements could be $\sim1.4^{-1/2}\approx0.8$ times
smaller than inferred assuming a $\Lambda$CDM model.

\subsection{Number counts of thermal Sunyaev-Zel'dovich clusters}

The most significant observable consequence of Early-DE models
identified so far is that the counts of cluster-sized halos should
decrease significantly more slowly than in $\Lambda$CDM
(cf.~Figs.~\ref{fig:6} and \ref{fig:7a}). While statistically complete
galaxy-cluster samples are notoriously hard to identify, future
cluster surveys based on the thermal Sunyaev-Zel'dovich (SZ) effect
\citep{SU72.1} should provide ideally suitable data for testing this
prediction of Early-DE models.

The upcoming \emph{Planck} satellite, due for launch in 2007, will
observe the entire microwave sky in frequency bands between 30 and
857~GHz with an angular resolution of down to 5 arc minutes. One of
its frequency bands is centred on 217~GHz where the (non-relativistic)
thermal Sunyaev-Zel'dovich effect vanishes. \emph{Planck}'s frequency
bands are thus well-placed for identifying the unique spectral
signature of the thermal SZ effect which reduces the CMB intensity
below, and increases it above, 217~GHz. \emph{Planck}'s high
sensitivity and comparatively high angular resolution will enable it
to detect a huge sample of clusters reaching high redshift.

\begin{figure}[ht]
  \includegraphics[width=\hsize]{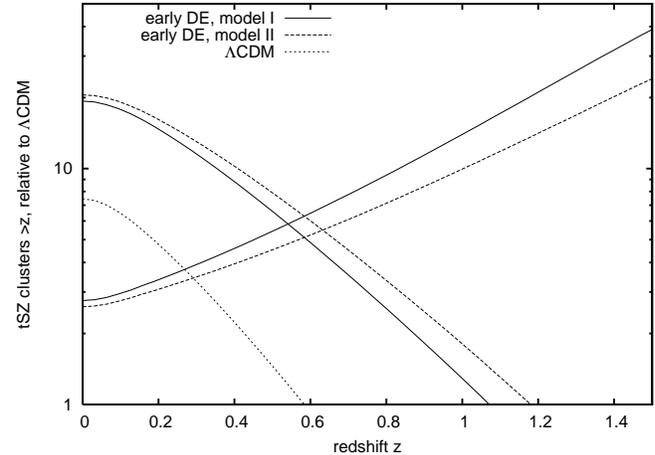}
\caption{\emph{Falling curves}: Cumulative redshift distribution of
  clusters expected to be detected through their thermal
  Sunyaev-Zel'dovich effect by the \emph{Planck} satellite. Curves are
  shown for the two models of Early Dark Energy and for $\Lambda$CDM,
  and they are normalised to show the number of clusters per 10 square
  degrees. Thermal-SZ cluster counts in the Early-DE models start at
  $z=0$ at approximately twice the value in $\Lambda$CDM and reach to
  substantially higher redshifts. The \emph{rising curves} show the
  redshift distributions in the Early-DE models, divided by the
  expectation in the $\Lambda$CDM model. For $z\ge0$, the total number
  of detections should increase by a factor of 2--3, while the
  number of clusters at redshifts $\ge1$ should be about an order of
  magnitude higher in Early-DE compared to $\Lambda$CDM models.}
\label{fig:11}
\end{figure}

Analytic estimates raise the expectation that \emph{Planck} may find
of order 30,000 galaxy clusters on the full sky \citep{SI00.1,BA03.1},
while simulations taking realistic foreground contamination, noise
patterns and full-sky, multi-band filtering techniques into account
arrive at lower numbers, $\lesssim10,000$ \citep{SA04.1}. Nonetheless,
the cluster sample expected from \emph{Planck} will be
enormous. Following the analytic description of the \emph{Planck}
cluster sample given in \cite{BA03.1}, we show in Fig.~\ref{fig:11}
the cumulative redshift distribution of clusters in the sample
expected from \emph{Planck} in the Early-DE models, normalised by the
expectation in the $\Lambda$CDM model.

While the total number of detections is moderately increased by a
factor of two to three, the number of high-redshift clusters
detectable in Early-DE models is substantially higher than in
$\Lambda$CDM. The number of clusters above redshift unity is already
an order of magnitude larger. Thus, the number of clusters detected by
\emph{Planck} will provide a highly sensitive test for Early-DE
compared to $\Lambda$CDM models.

Interestingly, observations of the CMB on small angular scales
($\lesssim5'$) with the \emph{Cosmic Background Imager} (CBI) indicate
a much higher temperature fluctuation amplitude than expected in
$\Lambda$CDM with the normalisation $\sigma_8\sim0.8$ typically
inferred from other observations \citep{PA01.1,MA03.4}. If due to the
thermal SZ effect, these measurements require $\sigma_8\gtrsim1$ in
the $\Lambda$CDM model \citep{BO05.1}, which would stretch other
normalisation constraints to their limits. Cosmological models with
Early Dark Energy such as those used here as examples could naturally
explain this ``CBI anomaly''.

\section{Summary and Conclusions}

We have investigated expectations for the growth of non-linear
structures in a model universe with Early Dark Energy. Such models
assume that the accelerated cosmic expansion is driven by a scalar
field which has non-negligible energy density all through cosmic
history. We have selected two examples for such models with Early Dark
Energy whose parameters are chosen such as to comply with the
temperature fluctuations measured in the cosmic microwave background.

We have generalised the spherical-collapse model such that Dark Energy
with variable equation-of-state parameter $w$ can be taken into
account. In our models the Dark Energy follows for $z\gtrsim1$ the
density of the dominant component in the mixture of cosmic
fluids. Modified in this way, the spherical collapse model predicts
that the linear density contrast necessary for collapse,
$\delta_\mathrm{c}$, is lowered compared to $\Lambda$CDM models or
Dark-Energy models with constant $w$.

This first result, which may appear unexpected, can be understood as
follows. The requirement of having virialised halos at a given
redshift sets the initial overdensity inside the spherical
perturbation idealising the later halo. Early Dark Energy reduces the
linear growth factor compared to $\Lambda$CDM due to its higher
expansion rate in the young universe. The initial overdensity required
for later collapse is thus extrapolated to lower linear overdensity at
collapse time. Lower linear density contrast is thus sufficient for
non-linear collapse, which increases the number of fluctuations
capable of forming halos.

The effects of Early Dark Energy on the geometrical properties of the
Universe are moderate. Assuming the same Hubble constant today, such
models are $\sim5\%$ younger than comparable $\Lambda$CDM models
today, and the angular-diameter distances are lowered by no more than
$\sim10\%$ even to high redshifts. Yet, these effects are noticeably
stronger than in Dark-Energy models with constant $w$; typically, they
are roughly twice as large than in a model universe which has
$w=-0.8$.

Since halos tend to grow earlier in Dark-Energy compared to
$\Lambda$CDM models, their core densities are higher, and so are their
concentration parameters. Using the prescription of Eke et al. for
computing expected halo concentrations, we found that halos at
redshift zero are more concentrated than in $\Lambda$CDM by $\sim20\%$
for galaxy masses, and by $\sim15\%$ for cluster masses. This increase
is visible, but smaller for halos forming at higher redshift.

Strong gravitational lensing profits non-linearly from higher halo
concentrations. On the other hand, the effective lensing distance
(proportional to the inverse critical surface density for lensing) is
lowered by Early Dark Energy, albeit weakly. At redshifts which are
typical for strong-lensing galaxies or clusters, $\sim0.3-0.8$, say,
we found a reduction of the effective lensing distances of order
$\sim8\%$ compared to $\Lambda$CDM. The power of weak gravitational
lensing by large-scale structures, however, is more substantially
changed. While there is no difference in the weak-lensing power
between Early Dark-Energy and $\Lambda$CDM models on large angular
scales $\gtrsim1^\circ$, it is increased by $\sim40\%$ on arc-minute
scales for sources at redshift $\sim1$. This implies a steeper
increase with decreasing angular scale of, e.g.~the two-point
correlation function of the cosmic shear than expected in
$\Lambda$CDM. It also implies that the normalisation of the
Dark-Matter power spectrum, $\sigma_8$, inferred from weak lensing
should be lowered by a factor of $\sim0.8$ as compared to its value
for a $\Lambda$CDM model.

The most pronounced effect, however, regards the present number
density of massive halos and its evolution towards higher redshift.
The Early Dark-Energy models we have studied here, normalised to the
CMB temperature-fluctuation measurements, predict approximately the
same number density of galaxy-sized halos today as expected in a
$\Lambda$CDM universe, and a number density of cluster-sized halos
which is $\sim40\%$ higher. Given the uncertainties in cluster counts
even at low redshift, this appears tolerable, although a very moderate
reduction of $\sigma_8$ would establish complete agreement at redshift
zero between the halo counts in Early Dark-Energy and $\Lambda$CDM
models. However, this present cluster population shrinks much less
quickly towards high redshift than in $\Lambda$CDM, implying that many
more of the clusters existing today were already present at redshifts
of order unity.

The cluster population above a mass limit of
$10^{14}\,h^{-1}\,M_\odot$ expected for $\sigma_8=0.8$ in Early
Dark-Energy models shrinks by a factor of $\sim2.5$ at redshift unity
compared to redshift zero, but by a factor of $\sim4$ in $\Lambda$CDM
models. This discrepancy increases rapidly for more massive
clusters. Raising the mass limit to $5\times10^{14}\,h^{-1}\,M_\odot$
makes the cluster population shrink by a factor of $\sim9$ in the
Early Dark-Energy models, but by a factor of $\sim28$ in an
equally-normalised $\Lambda$CDM model.

It appears reasonable here to distinguish two possibly different
values of $\sigma_8$ extracted from linear and non-linear structure
growth. As the mass function (\ref{eq:42}) shows, lowering
$\delta_\mathrm{c}$ due to Early Dark Energy can be compensated by
equally lowering $\sigma_8$. This implies that approximately the same
halo number at all redshifts as in the $\Lambda$CDM model can be
reproduced in the Early Dark Energy models by suitably lowering
$\sigma_8$. This $\sigma_8$ would then properly describe the abundance
of non-linear, collapsed structures. This can be tested against the
amplitude of linear structures, which is measurable for instance by
the galaxy power spectrum on large scales or weak gravitational
lensing on scales $\gtrsim10'$. In other words, if the value of
$\sigma_8$ extracted from non-linear structures within a $\Lambda$CDM
model turns out systematically higher than the $\sigma_8$ inferred
from larger structures, this would clearly hint at the presence of
Early Dark Energy.

For a given $\sigma_8$ measured from linear fluctuations, the Early
Dark Energy models predict a substantially slower evolution of the
halo population as in the $\Lambda$CDM model. This prediction has
several immediate observable consequences. First, the evolution of the
X-ray cluster luminosity function towards redshift unity should be
somewhat flatter than expected in a $\Lambda$CDM universe. Current
data are probably still insufficient for quantitative tests. Second,
the dynamical activity of clusters near redshift unity should be
substantially higher than expected in $\Lambda$CDM. The number of
major mergers experienced by cluster-sized halos near redshift unity
is predicted to be $\sim5$ times higher than in $\Lambda$CDM. Third,
and most prominently, the number of clusters expected to be found in
thermal Sunyaev-Zel'dovich surveys should increase
substantially. Applying a simple model for cluster detection by the
\emph{Planck} satellite, we found that $\sim2$ times more clusters
should be found at arbitrary redshifts by \emph{Planck} than in a
$\Lambda$CDM universe, but $\sim10$ times more above redshift
unity. Leaving the question aside of how redshifts could be measured
for clusters detected exclusively by their thermal Sunyaev-Zel'dovich
effect, this prediction will be testable very soon. We note that this
consequence of Early Dark Energy may naturally explain the high
fluctuation amplitude of CMB temperature fluctuations on arc-minute
scales, the so-called ``CBI anomaly'' \citep{PA01.1,MA03.4,BO05.1}.

Given these results, we arrive at the main conclusion that models for
Early Dark Energy will leave a measurable imprint on weak-lensing
measurements and determinations of the distance-redshift relation
which is measured using type-Ia supernovae, but their most
distinguishing signature is the slow evolution of the cluster
abundance between redshifts $\sim1$ and zero. Structure formation
being hierarchical, this also implies a similar effect on the
abundance evolution of galaxy-sized halos at substantially higher
redshifts, $z\gtrsim2-3$.

\acknowledgements{We are grateful to E.~Thommes for helpful
discussions.}

\appendix

\section{The virial overdensity\label{app:1}}

The conventional calculation used in many papers (see,
e.g.~\citealt{LA91.1,KI96.1,WA98.1,WE03.1,HO05.1,MA05.2}) starts from
the Friedmann equation (\ref{eq:2}) for the second time derivative
$\ddot y$ of the perturbation radius and converts it to an energy
equation after multiplication with $\dot y$. This energy equation
contains the cosmological constant or Dark-Energy density in a term
which appears analogous to a potential-energy term. Energy
conservation is then used by equating the total potential energy of
the overdense sphere at turn-around to the sum of kinetic and
potential energies at virialisation, which can be expressed solely in
terms of the potential energy by means of the virial theorem.

Following this approach, assuming that the Dark Energy does not clump
and therefore not participate either in the virialisation of the
collapsing overdensity (see \cite{MA05.2} for a thorough discussion),
we calculate $\Delta_\mathrm{v}$ and confirm that it is only weakly
modified by Early Dark Energy compared to, e.g.~the $\Lambda$CDM
result.

\begin{figure}[ht]
  \includegraphics[width=\hsize]{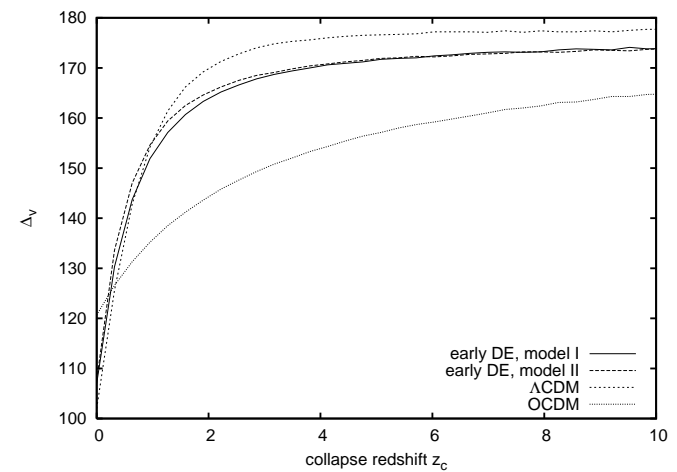}
\caption{The virial overdensity $\Delta_\mathrm{v}$ is shown as a
  function of collapse redshift $z_\mathrm{c}$ for the same four
  cosmologies used in Fig.~\ref{fig:2}. The change in
  $\Delta_\mathrm{v}$ due to Early Dark Energy is small compared to
  the difference between the more conventional models.}
\label{fig:1}
\end{figure}

We show in Fig.~\ref{fig:1} the virial overdensity $\Delta_\mathrm{v}$
calculated in this way as a function of the collapse redshift
$z_\mathrm{c}$ for two models of Early Dark Energy and compare it to
the results for a $\Lambda$CDM with $\Omega_\mathrm{m,0}=0.3$ and
$\Omega_\mathrm{\Lambda,0}=0.7$, and an OCDM model with
$\Omega_\mathrm{m,0}=0.3$. While $\Delta_\mathrm{v}$ is somewhat lower
at moderate and high $z_\mathrm{c}$ for the Early-DE models than for
$\Lambda$CDM, the difference between them remains moderate. For high
$z_\mathrm{c}$, they approach the ``canonical'' value of
$18\pi^2\approx178$ and fall towards $\Delta_\mathrm{v}\approx110$ for
$z_\mathrm{c}=0$. As shown for comparison, $\Delta_\mathrm{v}$ shows a
substantially flatter behaviour with $z_\mathrm{c}$ in the OCDM model.

We have, however, serious doubts about whether this approach is valid
in the context of cosmological models with (Early) Dark Energy. The
virial theorem concerns time-averaged energies of particles in bound
orbits. Dark-matter particles orbiting within collapsing
overdensities, however, will not feel any force from the Dark Energy
because the latter is homogeneously distributed. Thus, it appears
dubious to assign a potential to the Dark Energy whose gradient should
appear as a conservative force term in the equation of motion. Rather,
the Dark Energy should only appear acting on the expansion of the
background, and therefore as a dissipative force term which should
time-average out of the virial theorem. In this picture, the concept
of energy conservation also needs to be revisited. Since none of our
later results depends on $\Delta_\mathrm{v}$, we leave the discussion
at that point, showing $\Delta_\mathrm{v}$ as the conventional
approach predicts it for the Early Dark-Energy models, but expressing
qualms regarding the validity of the underlying physical concepts.

\end{document}